# Arch2030: A Vision of Computer Architecture Research over the Next 15 Years

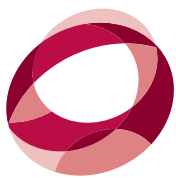

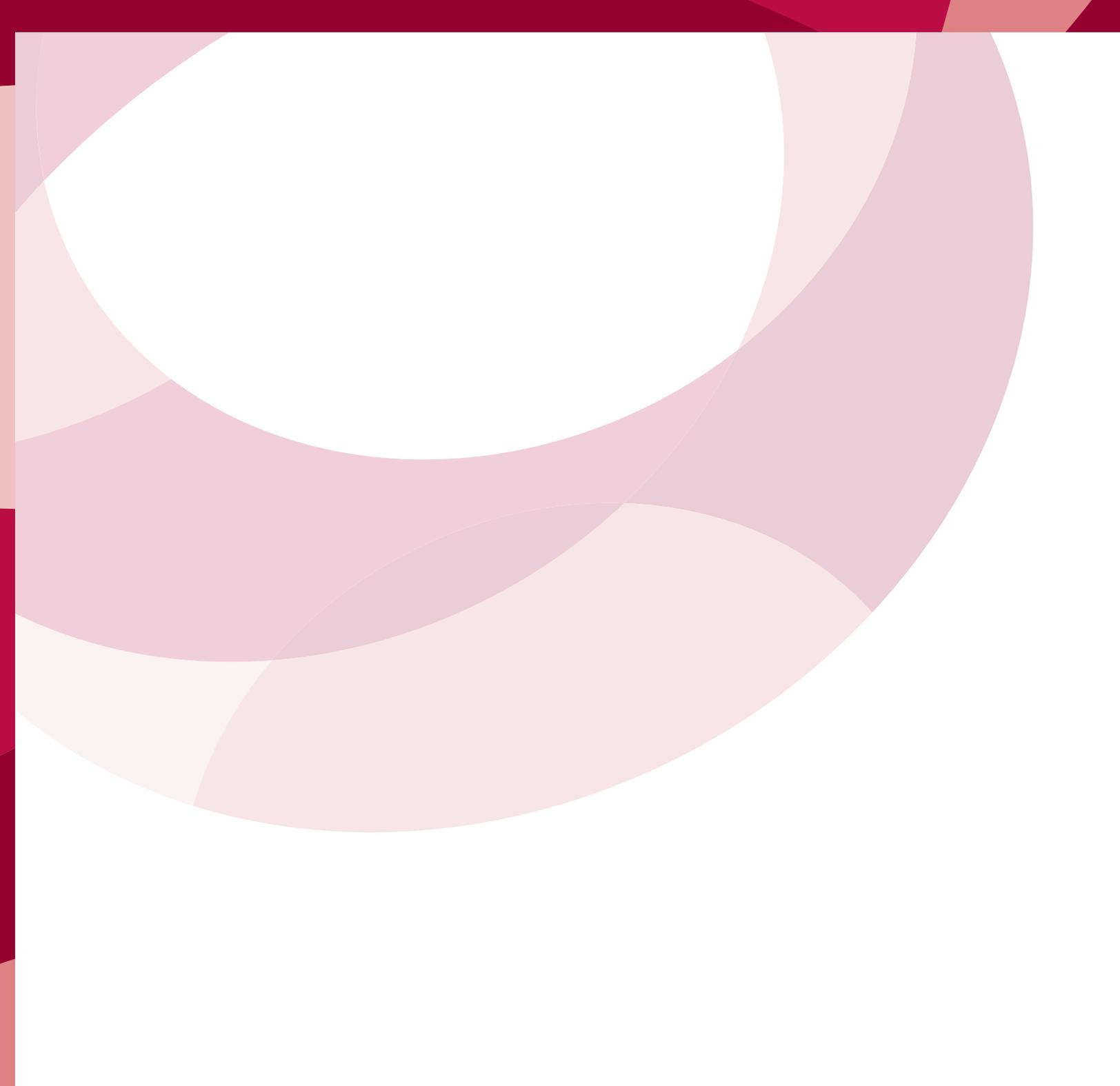

This material is based upon work supported by the National Science Foundation under Grant No. (1136993).

Any opinions, findings, and conclusions or recommendations expressed in this material are those of the author(s) and do not necessarily reflect the views of the National Science Foundation.

# Arch2030: A Vision of Computer Architecture Research over the Next 15 Years

Luis Ceze, Mark D. Hill, Thomas F. Wenisch

<>
Sponsored by

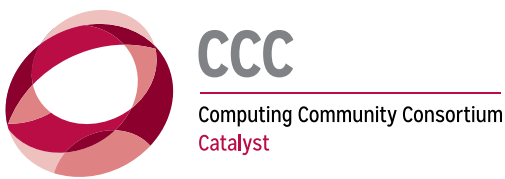

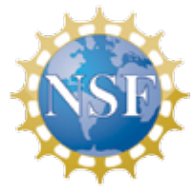

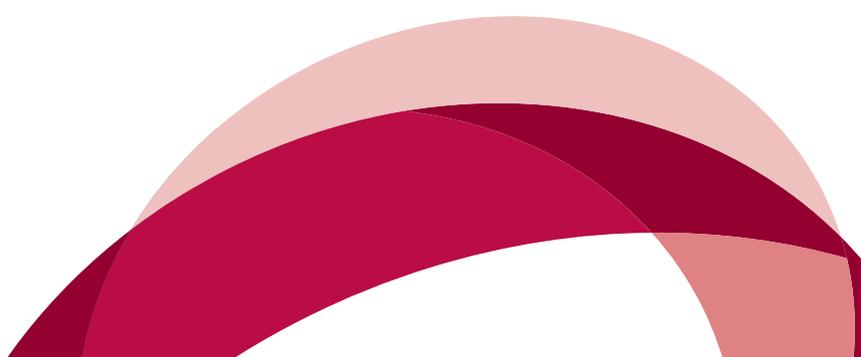

## Summary

Application trends, device technologies and the architecture of systems drive progress in information technologies. However, the former engines of such progress – Moore's Law and Dennard Scaling – are rapidly reaching the point of diminishing returns.  The time has come for the computing community to boldly confront a new challenge: how to secure a foundational future for information technology's continued progress.

The computer architecture community engaged in several visioning exercises over the years. Five years ago, we released a white paper, *21st Century Computer Architecture*, which influenced funding programs in both academia and industry. More recently, the *IEEE Rebooting Computing Initiative* explored the future of computing systems in the architecture, device, and circuit domains.

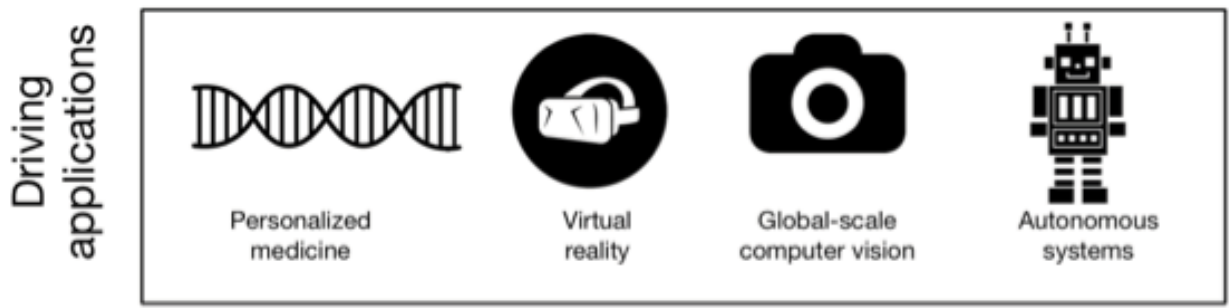

This report stems from an effort to continue this dialogue, reach out to the applications and devices/circuits communities, and understand their trends and vision. We aim to identify opportunities where architecture research can bridge the gap between the application and device domains.

Why now? A lot has changed in just five years:

1. We now have a clear **specialization gap** — a gap between off-the-shelf hardware trends and application needs. Some applications, like virtual reality and autonomous systems, cannot be implemented without specialized hardware, yet hardware design remains expensive and difficult.

2. **Cloud computing,** now truly ubiquitous, provides a clear "innovation abstraction;" the Cloud creates economies of scale that make ingenious, cross-layer optimizations cost-effective, yet offers these innovations, often transparently, to even the smallest of new ventures and startups.

3. **Going vertical** with 3D integration, both with die stacking and monolithic fabrication, is enabling silicon substrates to grow vertically, significantly reducing latency, increasing bandwidth, and delivering efficiencies in energy consumption.

4. **Getting closer to physics:** device and circuit researchers are exploring the use of innovative materials that can provide more efficient switching, denser arrangements, or new computing models, e.g., mixed-signal, carbon nanotubes, quantum-mechanical phenomena, and biopolymers.

5. And finally, **machine learning has emerged as a key workload;** in many respects, machine learning techniques, such as deep learning, caught system designers "by surprise" as an enabler for diverse applications, such as user preference prediction, computer vision, or autonomous navigation.

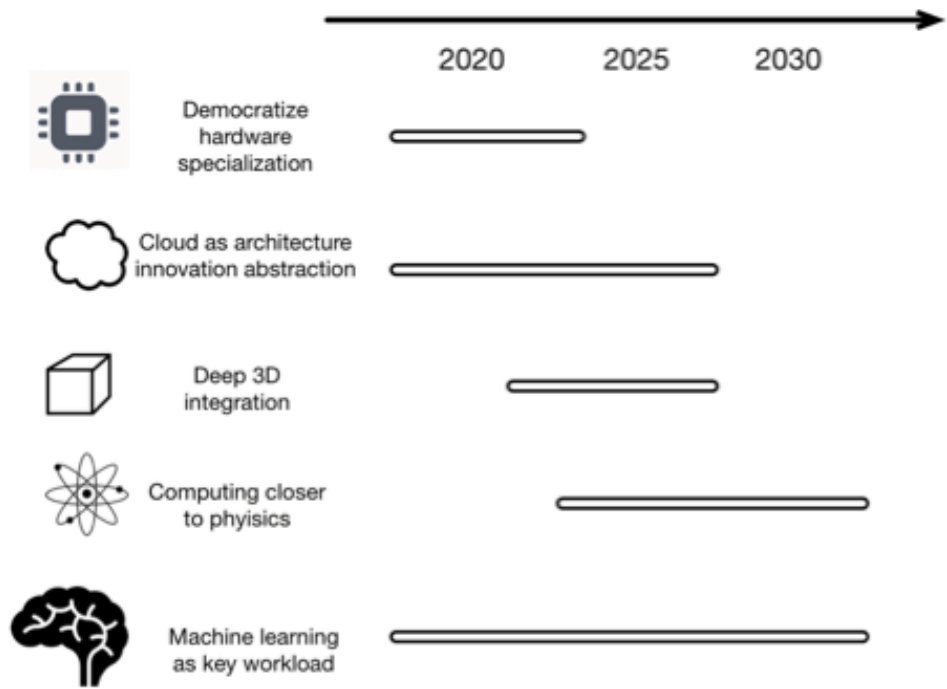

We now describe each opportunity in greater detail.

## The Specialization Gap: Democratizing Hardware Design

*Developing hardware must become as easy, inexpensive, and agile as developing software to continue the virtuous history of computer industry innovation.*

A widespread and emerging consensus maintains that classical CMOS technology scaling — the technical engine underlying Moore's Law that enables ever smaller transistors and denser integration — will come to an end in at most three more semiconductor technology generations (6-9 years)[1]. Further, Dennard scaling — the concomitant technical trend that enabled constant power per chip despite increasing CMOS integration density — ended in the mid-2000s[2,3], leading to a sea change in processor design: energy efficiency per operation has replaced area efficiency or peak switching speed as the most important design constraint limiting peak performance[4].

The effects of the imminent demise of classical scaling can be seen in recent industry announcements. Intel has abandoned its long-standing "tick-tock" model of releasing two major chip designs per technology generation, shifting instead to three designs; this extends the marketable lifetime of each generation as it drags the last gasps out of Moore's Law[5]. Further, the Semiconductor Industry Association has abandoned its biennial updates of the decades-old *International Technology Roadmap for Semiconductors*[6], a document that had been instrumental in coordinating technology, manufacturing, and system development across the industry. With no clear path to continued scaling, the value of the ITRS has ebbed.

---

[1] Chien and Karamcheti."Moore's Law: The First Ending and a New Beginning." *Computer* 46.12 (2013): 48-53.
[2] Fuller and Millett, "The Future of Computing Performance: Game Over or Next Level?," The National Academy Press, 2011 (http://books.nap.edu/openbook.php?record_id=12980&page=R1).
[3] Horowitz et al. "Scaling, power, and the future of CMOS." *IEEE International Electron Devices Meeting*, 2005.
[4] Mudge. "Power: A first-class architectural design constraint."*Computer* 34.4 (2001): 52-58.
[5] http://www.economist.com/technology-quarterly/2016-03-12/after-moores-law
[6] http://www.semiconductors.org/main/2015_international_technology_roadmap_for_semiconductors_itrs/

Yet, new applications continue to emerge that demand ever more computational capability. Foremost among these are the previously unimaginable applications enabled by large-scale machine learning, from image and speech recognition to self-driving cars to besting human experts at Go. Similar explosive growth can be seen in the need to process and understand visual data; some envisioned applications may demand the processing of gigapixels *per second for every human on earth*.

Past computing advances have been facilitated by the enormous investments in general-purpose computing designs enabled by classical scaling and made by only a handful of processor vendors. The large aggregate market of computing applications that benefited from these general-purpose designs amortized their substantial cost.

Given the twilight of classical scaling, continuing to meet emerging application performance demands by improving only a few general-purpose computing platforms is no longer feasible. Rather, over the past 5-10 years, a new strategy has emerged in some compute-intensive application domains: *specialized hardware design*. Specialized hardware (e.g., application-specific integrated circuits) can improve energy efficiency per operation by as much as 10,000 times over software running on a general-purpose chip[7]. The energy efficiency gains of specialization are critical to enable rich applications in the emerging Internet-of-Things. Specialization has been enormously successful in graphics rendering and video playback. Other initial evidence of commercial success is in machine learning applications. Indeed, the computer architecture research community has recognized and embraced specialization: of 175 papers in the 2016 computer architecture conferences (ISCA, HPCA, MICRO), 38 papers address specialization with GPUs or application-specific accelerators, while another 17 address specialized designs for machine learning.

However, commercial success of specialized designs, to date, has been demonstrated only for applications with enormous markets (e.g., video games, mobile video playback) that can justify investments of a scale similar to those made by general-purpose processor vendors. In terms of both time-to-market and dollars, the cost of designing and manufacturing specialized hardware is prohibitive for all but the few designs that can amortize it over such extensive markets.

To continue the virtuous innovation cycle, it is critical to reduce the barriers to application specific system design; to enable the energy efficiency advantages of specialization for all applications. Our vision is to "democratize" hardware design; that is, that hardware design become as agile, cheap, and open as software design. Software development teams can leverage a rich ecosystem of existing reusable components (often free and open source), use high-level languages to accelerate the capability of an individual developer, and rely on capable and automated program analysis, synthesis, testing, and debugging aids that help ensure high quality.

Despite decades of investment, computer-aided design has not delivered the same level of capability for hardware to a small development team. System designers require better tools to facilitate higher productivity in hardware description, more rapid performance evaluation, agile prototyping, and rigorous validation of hardware/software co-designed systems. Tool chains must mature to enable easy retargeting across multiple hardware substrates, from general purpose programmable cores to FPGAs, programmable accelerators, and ASICs. Better abstractions are needed for componentized/reusable hardware, possibly in the form of synthesizable intellectual property blocks or perhaps even physical chips/chiplets that can be integrated cheaply at manufacture. The architecture research community has an opportunity to lead in the effort to bridge the gap between general-purpose and specialized systems and deliver the tools and frameworks to make democratized hardware design a reality.

---

[7] Hameed et al. "Understanding sources of inefficiency in general-purpose chips." *International Symposium on Computer Architecture*, 2010.

## The Cloud as an Abstraction for Architecture Innovation

*By leveraging scale and virtualization, Cloud computing providers can offer hardware innovations transparently and at low cost to even the smallest of their customers.*

The disruptive nature of Cloud computing to business-as-usual has been widely appreciated[8]. The Cloud lets new ventures scale far faster than traditional infrastructure investment. New products can grow from hundreds to millions of users in mere days, as evidenced by the meteoric launch of Pokemon Go in July 2016. However, the Cloud also disrupts traditional Fortune 500 business models since businesses that previously owned their own IT infrastructure realize the cost benefits derivable from leasing Cloud resources.

Less widely appreciated, however, is the Cloud computing model's ability to provide a powerful abstraction for cross-layer architectural innovation that was previously possible in only a very few, vertically integrated IT sectors (e.g., specialized high-performance supercomputers). The model provides two critical advantages: *scale* and *virtualization*.

Cloud computing providers can leverage scale not only for their own businesses, but for the benefit of their customers making investments in IT. As a result, these providers often find it cost effective to make enormous, non-recurring engineering investments, for example, to develop entirely new hardware and software systems in-house rather than relying on third-party vendor offerings.

We are beginning to see the emergence of specialized computer architectures enabling unprecedented performance in the Cloud. GPUs are becoming ubiquitous, not only in high-end supercomputers, but also in commercial Cloud offerings. Microsoft has publicly disclosed Catapult[9], its effort to integrate field-programmable gate arrays to facilitate compute specialization in its data centers. Cavium has released the ThunderX, a specialized architecture for Internet service applications. Google has disclosed the Tensor Processing Unit[10], a dedicated co-processor for machine learning applications. These projects demonstrate that the economic incentives are in place for Cloud providers to invest in computer architecture specialization.

For academic computer architecture researchers, now is the moment to seize this opportunity and present compelling visions for cross-layer specialization. For example, the ASIC Clouds effort presents a vision for how a large number of highly specialized processors can be deployed in concert to drastically accelerate critical applications[11]. The scale of the Cloud computing landscape has created a viable path for such academic proposals to demonstrate real, immediate impact. Another aspect of in-house specialization is the use of technologies that require special facilities, for example, atomic clocks for global time synchronization or superconducting logic that requires extremely low temperatures and makes sense only in a data-center environment.

The second critical advantage of the Cloud computing model is *virtualization*. By virtualization, we refer to a broad class of techniques that introduce new hardware and software innovations *transparently* to existing software systems. Virtualization lets a Cloud provider swap out processing, storage, and networking components for faster and cheaper technologies without requiring coordination with their customers. It also enables the oversubscription of resources — transparent sharing among customers with time-varying, fractional needs for a particular resource. Oversubscription is essential to the cost structure of Cloud computing: it lets Cloud providers offer IT resources at far lower prices than those individual customers would incur by purchasing dedicated resources.

---

[8] http://www.zdnet.com/article/eight-ways-that-cloud-computing-will-change-business/

[9] Putnam, et al. "A reconfigurable fabric for accelerating large-scale datacenter services." *ACM/IEEE 41st International Symposium on Computer Architecture,* 2014.

[10] https://cloudplatform.googleblog.com/2016/05/Google-supercharges-machine-learning-tasks-with-custom-chip.html

[11] Magaki et al. "ASIC Clouds: Specializing the Datacenter." *ACM/IEEE 43rd International Symposium on Computer Architecture,* 2016.

Academic computer architecture research has long been fundamental to enabling virtualization; indeed, VMWare, the most recognizable vendor of virtualization technology, was launched from a university research project. Academic architecture researchers must continue to play a key role in developing virtualization techniques that close the gap between virtualized and bare-metal performance. And, architecture researchers must develop new virtualization abstractions to enable transparent use and oversubscription of specialized hardware units, like the Catapult, TPU, or ASIC clouds.

## *Going Vertical*

*3D integration provides a new dimension of scalability.*

A critical consequence of the end of Moore's Law is that chip designers can no longer scale the number of transistors in their designs "for free" every 18 months. Furthermore, over recent Silicon generations, driving global wires has grown increasingly expensive relative to computation, and hence interconnect accounts for an increasing fraction of the total chip power budget.

3D integration offers a new dimension of scalability in chip design, enabling the integration of more transistors in a single system despite an end of Moore's Law, shortening interconnects by routing in three dimensions, and facilitating the tight integration of heterogeneous manufacturing technologies. As a result, 3D integration enables greater energy efficiency, higher bandwidth, and lower latency between system components inside the 3D structure.

Architecturally, 3D integration also implies that computing must be near data for a balanced system. While 3D has long enabled capacity scaling in Flash and other memory devices, we are only now beginning to see integration of memory devices and high performance logic, for example, in Micron's Hybrid Memory Cube. 3D stacking has prompted a resurgence of academic research in "near-data computing" and "processing-in-memory" architectures, because it enables dense integration of fast logic and dense memory. Although this research topic was quite popular 20 years ago, processing-in-memory saw no commercial uptake in the 1990s due to manufacturability challenges. With the advent of practical die stacking and multi-technology vertical integration, such architectures now present a compelling path to scalability.

While 3D integration enables new capabilities, it also raises complex new challenges for achieving high reliability and yield that can be addressed with architecture support. For example, 3D-integrated memory calls to re-think traditional memory and storage hierarchies. 3D integration also poses novel problems for power and thermal management since traditional heat sink technology may be insufficient for the power density of high-performance integrated designs. Such problems and challenges open a new, rich field of architectural possibilities.

## *Architectures "Closer to Physics"*

*The end of classical scaling invites more radical changes to the computing substrate.*

New device technologies and circuit design techniques have historically motivated new architectures. Going forward, several possibilities have significant architectural implications. These fall into two broad categories. The first is *better use of current materials and devices* by a more efficient encoding of information, one closer to analog. There has been a rebirth of interest in analog computing because of its good match to applications amenable to accuracy trade-offs. Further, analog information processing offers the promise of much lower power by denser mapping of information into signals and much more efficient functional units than their digital counterparts. However, such computing, more subject to noise, requires new approaches to error tolerance for it to make sense.

The second category of opportunities is *the use of "new" materials,* which can cover more efficient switching, denser arrangements, and unique computing models. Below we list a few prominent efforts worthy of the architecture community's attention.

**New memory devices.** For decades, data has been stored in DRAM, on Flash, or on rotating disk. However, we are now on the cusp of commercial availability

ARCH2030: A VISION OF COMPUTER ARCHITECTURE RESEARCH OVER THE NEXT 15 YEARSof new memory devices (e.g., Intel/Micron 3D XPoint memory[13]) that offer fundamentally different cost, density, latency, throughput, reliability, and endurance trade-offs than traditional memory/storage hierarchy components.

**Carbon nanotubes.** Electronics based on carbon nanotubes (CNTs) continues to make significant progress, with recent results showing simple microprocessors implemented entirely with CNTs[14]. CNTs promise greater density and lower power and can also be used in 3D substrates. This momentum makes CNTs a viable area for architects' consideration.

**Quantum computing.** Quantum computing uses quantum mechanics phenomena to store and manipulate information. Its key advantage is that the "superposition" quantum phenomenon effectively allows representation of 0 and 1 states simultaneously, which can be leveraged for exponential speed-ups compared to classical computing for select algorithms.

A sister effort of quantum computing is **superconducting logic.** Systems that use superconducting devices, such as Josephson junctions, offer "free" communication because they consume little energy to move a signal over a superconducting wire[12]. Operations on data, on the other hand, are more expensive than moving data. These trade-offs are the reverse of those in silicon CMOS, where most energy is dissipated in communication rather than operations on the data path.

Microsoft, Google, IBM and I-ARPA have publicized significant investments in quantum computing and superconducting logic. We conclude that the time is ripe for renewed academic interest in quantum computer architectures, with a likely path to practical impact within a decade.

**Borrowing from biology.** The use of biological substrates in computing has long been considered a possibility in several aspects of computer systems. DNA computing has demonstrated simple logic operations and more recent results show the potential of using DNA as a digital medium for archival storage and for self-assembly of nanoscale structure[15]. Progress in DNA manipulation[16] fueled by the biotech industry is making the use of biomaterials a more viable area for consideration among architecture researchers. Beyond DNA, there are other biomolecules that could be used for computing such as proteins, whose engineering advanced significantly in the past decade[17].

## *Machine Learning as a Key Workload*

*Machine Learning is changing the way we implement applications. Hardware advancement makes machine learning over big data possible.*

Machine learning (ML) has made significant progress over the last decade in producing applications that have long been in the realm of science fiction, from long-sought, practical voice-based interfaces to self-driving cars. One can claim that this progress has been largely fueled by abundant data coupled with copious compute power. Large-scale machine learning applications have motivated designs that range from storage systems to specialized hardware (GPUs, TPUs).

While the current focus is on supporting ML in the Cloud, significant opportunities exist to support ML applications in low-power devices, such as smartphones or ultra-low power sensor nodes. Luckily, many ML kernels have relatively regular structures and are amenable to accuracy-resource trade-offs; hence, they lend themselves to hardware specialization, reconfiguration, and approximation techniques, opening up a significant space for architectural innovation.

---

[12] "Superconducting Computing and the IARPA C3 Program", http://beyondcmos.ornl.gov/documents/Session%203_talk1-Holmes.pdf

[13] http://www.intel.com/content/www/us/en/architecture-and-technology/non-volatile-memory.html

[14] https://www.technologyreview.com/s/519421/the-first-carbon-nanotube-computer/

[15] http://people.ee.duke.edu/~dwyer/pubs/TVLSI_dnaguided.pdf

[16] http://www.synthesis.cc/synthesis/2016/03/on_dna_and_transistors

[17] http://www.sciencemag.org/news/2016/07/protein-designer-aims-revolutionize-medicines-and-materials

Machine learning practitioners spend considerable time on computation to train their models. Anecdotal evidence suggests that week- to month-long training jobs are common, even when using warehouse-scale infrastructure. While such computational investments hopefully amortize over many invocations of the resulting model, the slow turnaround of new models can negatively affect the user experience. Consequently, architecture researchers have new opportunities to design systems that better support ML model training.

## About this document

This document resulted from discussions held during the Arch2030 Workshop[18] at ISCA 2016, organized by Luis Ceze and Thomas Wenisch and shepherded by Mark Hill. The organizers also solicited input from the community in the form of an open survey as well as direct comments on this document. Ceze and Wenisch drafted the text and revised it based on community feedback. The contributors listed below provided feedback on drafts of this document and have granted their endorsement to its content.

Endorsers:

Luis Ceze, University of Washington
Thomas F. Wenisch, University of Michigan
Mark D. Hill, University of Wisconsin-Madison
Sarita Adve, University of Illinois at Urbana-Champaign
Alvin R. Lebeck Duke University
Michael Taylor, University of California San Diego
Josep Torrellas, University of Illinois at Urbana Champaign
Karin Strauss, Microsoft
Dan Sorin, Duke University
Doug Burger, Microsoft
Tom Conte, Georgia Institute of Technology, co-chair of the IEEE Rebooting Computing Initiative
Babak Falsafi, EPFL
Fred Chong, University of Chicago
Rakesh Kumar, University of Illinois at Urbana-Champaign
Todd Austin, University of Michigan
Christos Kozyrakis, Stanford University
Karu Sankaralingam, UW-Madison
James Tuck, NC State University
Trevor Mudge, University of Michigan
Martha Kim, Columbia University
Stephen W. Keckler, NVIDIA
Vikram Adve, University of Illinois at Urbana-Champaign
Timothy Sherwood, UC Santa Barbara
Kathryn S McKinley, Microsoft Research
Yuan Xie, UCSB
Lieven Eeckhout, Ghent University
Andrew Putnam, Microsoft
Nikos Hardavellas, Northwestern University
James Larus, EPFL IC
Simha Sethumadhavan, Columbia University
Andreas Moshovos, University of Toronto
David J. Lilja, University of Minnesota
Guri Sohi, University of Wisconsin-Madison
Antonio Gonzalez, UPC Barcelona
Jack Sampson, Penn State
Natalie Enright Jerger, University of Toronto
Mark Oskin, University of Washington
Ulya Karpuzcu, University of Minnesota
David Kaeli, Northeastern University

---

[18] The workshop was supported by the Computing Community Consortium.

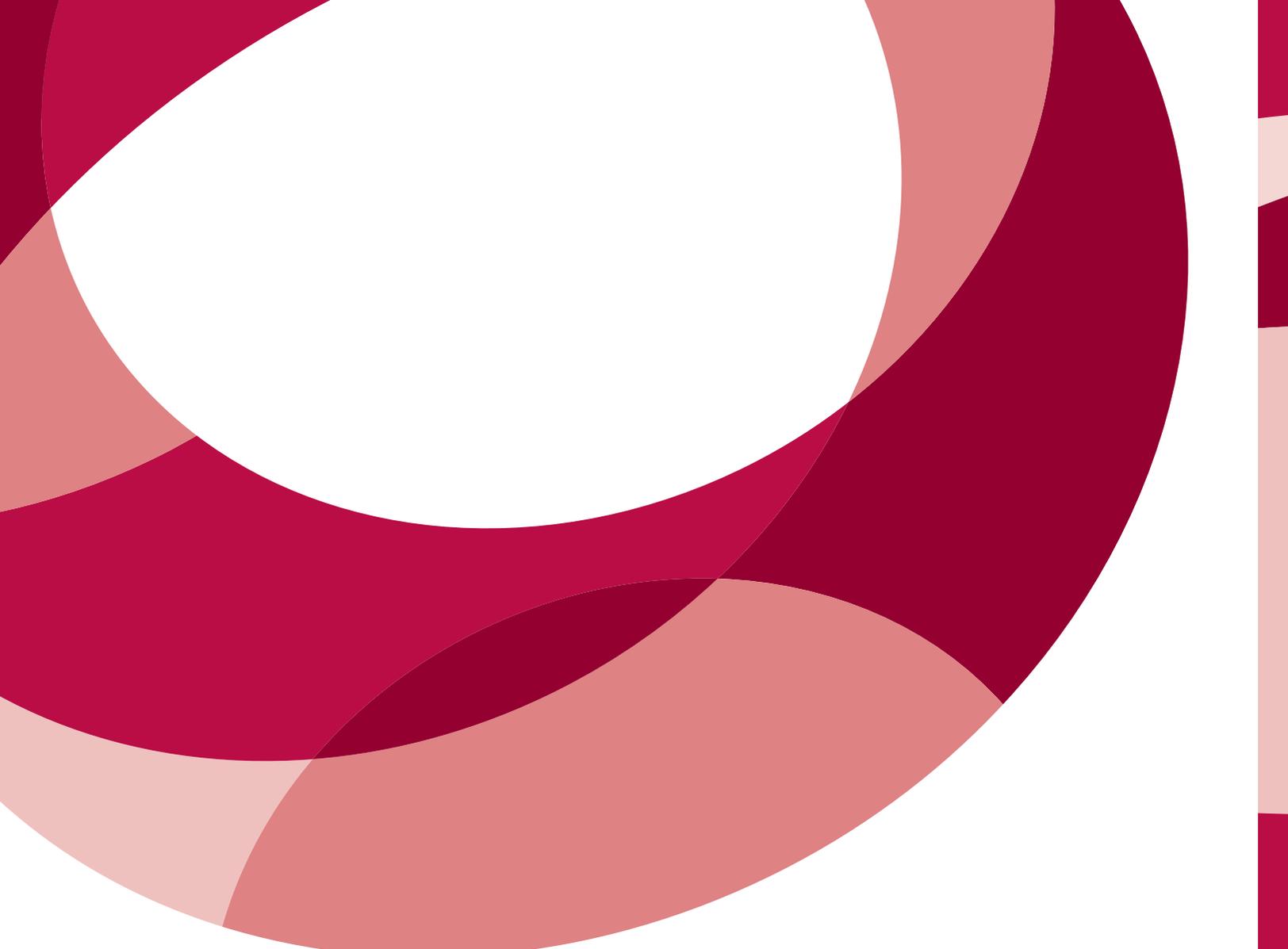

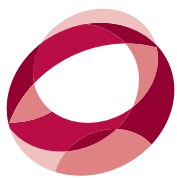

**CCC**
Computing Community Consortium
Catalyst

1828 L Street, NW, Suite 800
Washington, DC 20036
P: 202 234 2111 F: 202 667 1066
www.cra.org cccinfo@cra.org